\documentclass[12pt]{article}
\usepackage[dvips]{graphicx}
\usepackage{pdproc}

  \makeatletter 
  \def\@cite#1{[#1]} 
  \makeatother    
\begin{document}

\renewcommand{\thefootnote}{\alph{footnote}}

%%%%%%%%%%%%%%%%%%%%%%%%%%%%%%%%
% beginning of our definitions
%%%%%%%%%%%%%%%%%%%%%%%%%%%%%%%%

\newcommand{\lsim}{\raisebox{-0.13cm}{~\shortstack{$<$ \\[-0.07cm] $\sim$}}~}
\newcommand{\Raw}{\Rightarrow}
\newcommand{\MNS}{{\cal U}}

\vspace*{-20mm}
\begin{flushright}
 CAFPE-37/04 \\
UG-FT-167/04 \\ \
\end{flushright}

%%%%%%%%%%%%%%%%%%%%%%%%%%%%%%%%
% end of our definitions
%%%%%%%%%%%%%%%%%%%%%%%%%%%%%%%%

\title{
       Mini-review on Lepton Flavor and CP Violation in SUSY\footnote{
Talk presented by J.I.I. at the 12th International Conference on Suypersymmetry
and Unification of the Fundamental Interactions, June 17-23, 2004, in
Epochal Tsukuba, Tsukuba, Japan.}
}

\author{JOS\'E I. ILLANA and MANUEL MASIP}

\address{
Centro Andaluz de F{\'\i}sica de Part{\'\i}culas Elementales (CAFPE) and \\ 
Departamento de F{\'\i}sica Te\'orica y del Cosmos, Universidad de Granada, \\
E-18071 Granada, Spain
%%%%% You may comment out the e-mail address line below.  
\\ {\rm E-mails: jillana@ugr.es, masip@ugr.es}}

\abstract{
Lepton flavor and CP violation in supersymmetric models are briefly reviewed.
After a short motivation and an introduction to the phenomenology, model
independent constraints on mass insertions, predictions of SUSY GUT models 
and rates for several LFV and CPV processes are presented.
}

\normalsize\baselineskip=15pt

\section{Motivation}

In spite of the impressive success of the standard model (SM), there are 
two pieces of evidence for physics beyond, both related to the
subject of this talk: lepton flavor violation (LFV), CP violation and
supersymmetry (SUSY). 
The first, neutrino oscillations (neutral LFV), indicate non-zero 
neutrino masses and mixings (their size explained naturally 
by the seesaw mechanism). 
The second, the observed baryon asymmetry of the Universe, 
requires new sources of CP violation (leptogenesis could
be induced by heavy Majorana neutrinos, and electroweak baryogenesis 
is possible in SUSY extensions of the SM).

Neutrino masses introduce negligible charged LFV effects (one loop, 
GIM suppressed) and negligible CPV effects (three loops) in the SM. 
In contrast, embedded in a SUSY framework the flavor structure that
they suggest implies new sources of FV (misalignment between lepton 
and slepton mass matrices) and CP violation that show up to one loop 
and are GIM unsuppressed. In the seesaw mechanism SUSY would be
necessary to cancel the large corrections to the Higgs mass 
proportional to the Majorana scale.

\section{Phenomenology}

In the minimal supersymmetric SM (MSSM) with R-parity and 
seesaw neutrino mas\-ses,
the relevant parameters are the heavy Majorana neutrino masses,
the complex parameter $\mu$, charged
lepton and neutrino Yukawa couplings (in the superpotential) and complex 
SUSY soft-breaking bilinear and trilinear terms (flavor-diagonal gaugino
masses and flavor-nondiagonal, CP-violating slepton mass matrices). At the
electroweak scale heavy neutrinos decouple and our observables are the
SUSY masses and mixings (bounded by LFV decays and EDMs of charged leptons)
and the light neutrino masses and mixings (partially measured in neutrino 
oscillation experiments). The neutrino Yukawa matrix ${\bf Y_\nu}$, in the 
basis where the heavy-neutrino mass matrix ${\bf M_N}$ is ${\bf D_M}$ diagonal,
can be obtained only up to a complex orthogonal matrix ${\bf R}$
from the light-neutrino (symmetric) mass matrix 
${\cal M}_\nu$ (diagonalized by ${\bf D_m}=\MNS^T{\cal M}_\nu\ \MNS$)
\cite{Casas:2001sr}:
\begin{equation}
{\cal M}_\nu=v^2\sin^2\beta {\bf Y}^T_\nu {\bf M^{\rm -1}_N} {\bf Y_\nu}
\quad\Raw\quad
v\sin\beta\ {\bf Y_\nu}=\sqrt{{\bf D_M}}{\bf R}\sqrt{{\bf D_m}}
\ \MNS^\dagger\ .
\end{equation}
A redefinition of the fields and the symmetries of the Lagrangian lead to
a few physical phases: one Dirac and two Majorana phases in the MNS matrix
(${\cal U}$ in the basis where ${\bf Y_e}$ is diagonal), the SUSY phase
$\arg(\mu)$ (in chargino and neutralino mass matrices), the SUSY-breaking
phase $\arg(A_0)$ (the trilinear coupling, taken universal at some scale, 
in the left-right mixing of the slepton mass matrix) and the three phases 
of ${\bf R}$.

The relevant interaction vertices of the mass eigenstates 
$\tilde\nu^\dagger_X\tilde\chi^-_A\ell_I$ and
$\tilde\ell^\dagger_X\tilde\chi^0_A\ell_I$ can be parametrized by 
$ig(c^L_{IAX}P_L+c^R_{IAX}P_R)$ where $P_{L(R)}=\frac{1}{2}(1\mp\gamma_5)$
and the couplings $c^{L,R}_{IAX}$ involve chargino, neutralino, slepton 
and lepton mixings. At present the most stringent constraints on these 
couplings come from LFV decays $\ell_J\to\ell_I\gamma$, the muon $(g-2)$ 
and the electron EDM which are all (chirality-flipping) dipole 
transitions/moments  that can be 
parametrized by magnetic and electric dipole form factors. Charginos-sneutrinos
and neutralinos-charged sleptons are exchanged in one-loop triangular diagrams.
The form factors in terms of both generic couplings and SUSY parameters can 
be found in Ref.~\cite{Illana:2003pj}.

\section{Model-independent constraints}

The $(6\times6)$ charged slepton and $(3\times3)$ sneutrino mass matrices can
 be rewritten in terms of several types of mass insertions (flavor basis
$\alpha,\beta=1,2,3$) \cite{insertions}:
\begin{equation}
(\delta^{\tilde\nu}_{LL})_{\alpha\beta}\equiv
\frac{(M^2_{\tilde\nu})_{\alpha\beta}}{\tilde m^2},\
(\delta^{\tilde\ell}_{LL})_{\alpha\beta}\equiv\frac{(m^2_{LL})_{\alpha\beta}}
{\tilde m^2},\
(\delta^{\tilde\ell}_{RR})_{\alpha\beta}\equiv\frac{(m^2_{RR})_{\alpha\beta}}
{\tilde m^2},\
(\delta^{\tilde\ell}_{LR})_{\alpha\beta}\equiv\frac{(m^2_{LR})_{\alpha\beta}}
{\tilde m^2},
\end{equation}
where $\tilde m$ is an average slepton mass.
This parametrization is (SUSY) model independent. The flavor non-diagonal 
mass insertions ($\alpha\ne\beta$) are constrained experimentally in LFV 
processes, assuming that only one type of deltas contribute to a given
process, ignoring potential cancellations.
Notice that $\delta_{LL}$ and $\delta_{RR}$ are hermitian, so that
${\rm Im}(\delta_{LL,RR})_{\alpha\alpha}=0$. 
This means that the only contribution from the chargino-sneutrino sector
to the EDM comes from the SUSY phase $\arg(\mu)$.

The form factors are dominated by contributions proportional to
$c^{L(R)^*}_{IAX}c^{R(L)}_{JAX}$ (the chirality flip takes place in internal 
lines) and go like
$m_{\ell_J}\tan\beta\,\delta^{\tilde\nu}_{LL}$,
$m_{\ell_J}\tan\beta\,\delta^{\tilde\ell}_{LL,RR}$,
$M_1\delta^{\tilde\ell}_{LR}$.
This means that bounds on $\delta^{\tilde\ell}_{LR}$ are essentially 
independent of $\tan\beta$ and stronger than the others, which in contrast 
are more restrictive for large $\tan\beta$. The $\delta_{LL}$ insertions 
receive both U(1) and SI(2) contributions. There are two dominant U(1) 
contributions for large $\tan\beta$ to $\delta_{LL(RR)}$ that have 
same (opposite) signs. The cancellations make $\delta_{RR}$ be 
less constrained than $\delta_{LL}$ \cite{Hisano+Masina}. 
The branching ratios roughly grow like $\delta^2_{LL,RR}\tan^2\beta$.

The present limit BR$(\mu\to e\gamma)<1.2\times10^{-11}$ strongly constrains
the LFV mass insertions between the first two families in the charged slepton
sector, except for incidental cancellations of neutralino contributions, 
leading to $(\delta^{\tilde\ell}_{LL,RR})_{12}\lsim10^{-5}$ to 
$10^{-3}$ (stronger for higher $\tan\beta$) and 
$(\delta^{\tilde\ell}_{LR})_{12}\lsim10^{-6}$. 
Due to the large lepton mixing (neutrino oscillations) the same 
limit constrains all the sneutrino mass insertions 
$(\delta^{\tilde\nu}_{LL})_{12,13,23}\lsim10^{-5}$ to 
$10^{-3}$. Since BR$(\tau\to e\gamma,\,\mu\gamma)\lsim10^{-7}$,
the LFV mass insertions involving the third family are little constrained.
See for instance Refs.~\cite{Illana:2002tg,Branco:2003zy}.

The present limit on the electron EDM $d_e\lsim10^{-27}\ e$cm restricts the
imaginary part of the flavor-diagonal insertion 
$(\delta^{\tilde\ell}_{LR})_{ee}$ (to the level of $10^{-8}$) and 
$\arg(\mu)$ (less strongly due to a possible conspiracy of Majorana phases). 
From present limits on $d_\mu$ and $d_\tau$ there are no significant 
constraints. See Ref.~\cite{Branco:2003zy}.

%Correlation among AMDM, EDM and LFV.

\section{Predictions for radiative lepton decays in SUSY GUT models}

The experimental limits imply a very fine alignment of lepton and slepton
mass matrices. This is generally formulated as the SUSY flavor problem: how 
to explain such an alignment when fermion and sfermion masses have a
different origin, the former from Yukawa 
interactions and the latter from a soft SUSY-breaking mechanism.

Even if the SUSY-breaking mass matrix was diagonal and universal, as in SUGRA, 
at a high scale, the renormalization group (radiative) corrections down to 
the electroweak scale introduce off-diagonal entries due to non universal 
Yukawa interactions. 

Assuming universality at the GUT scale $M_X$, there are corrections 
proportional to $({\bf Y_\nu^\dagger Y_\nu})_{ij}\log(M_X/M)$ in the 
slepton left-handed sector. Since $Y_\nu\sim\sqrt{m_\nu M}/(v\sin\beta)$, 
LFV is enhanced for $M\sim10^{14}$~GeV ($Y_\nu\sim1$). The scenario with 
minimal LFV is that of quasidegenerate neutrinos and a 
real matrix ${\bf R}$ (though incompatible with leptogenesis 
\cite{Pascoli:2003rq}). These corrections are then independent of ${\bf R}$. 
One can distinguish two cases compatible with the lepton mixings observed at 
low energy \cite{Illana:2003pj}: 
(a) high $\tan\beta$ ($Y_\tau\sim1$) with CKM-like mixings at GUT scale
(radiative magnification \cite{radmagnif}) and 
(b) low $\tan\beta$ ($Y_\tau\ll1$) where lepton mixing angles do not run. 
Assuming universality at the Planck scale $M_P$, in the simplest SU(5)
grand unification model, there are also top-quark induced corrections 
proportional to $({\bf Y_u^\dagger Y_u})_{ij}\log(M_P/M_X)$ in the 
slepton right-handed sector. Present limits
on $\mu\to e\gamma$ exclude case `a' for large neutrino couplings,
and the non observation in the future of $\mu\to e\gamma$ at PSI or
$\tau\to\mu\gamma$ at KEK would exclude the magnification model 
\cite{Illana:2003pj}.

Quarks and leptons share multiplets in GUT models, leading to a
correlation between leptonic and hadronic FCNC and CP \cite{oscar}. 
In SU(5), BR$(\tau\to\mu\gamma)$ restricts the SUSY contribution to
$A_{CP}(B\to\phi K_S)$ and limits on $(\delta^{\tilde d}_{RR})_{12(13)}
\leftrightarrow(\delta^{\tilde \ell}_{LL})_{12(13)}$
from $\Delta m_K$ and $B_0-\bar B_0$ compete with those from
BR$(\mu\to e\gamma)$ and BR$(\tau\to e\gamma)$, respectively.
In SO(10), limits on BR$(\ell_J\to\ell_I\gamma)$ complement SUSY
direct searches at LHC, and $\tau\to\mu\gamma$ is more restrictive
than $b\to s\gamma$.

\section{Other LFV processes}

$\ell_J\to 3\ell_I$: The contribution of photon-exchange penguins to the 
branching ratio is ${\cal O}(\alpha)$ smaller than the 
BR$(\ell_J\to\ell_I\gamma)$ ($Z$-penguins and box diagrams are subdominant)
and that of Higgs-penguins grow with $\tan^6\beta/M^2_A$ \cite{Babu:2002et}. 
In particular, limits from BR$(\tau\to3\mu)$ compete with those from 
BR$(\tau\to\mu\gamma)$ for light Higgs masses and large $\tan\beta$.

$\mu\to e$ conversion in nuclei: The contribution of photon-penguins to 
the conversion rate is $\sim 6\times10^{-3}$ smaller than the 
BR$(\mu\to e\gamma)$, but Higgs-penguin contributions 
$\propto\tan^6\beta/M^2_A$ may dominate \cite{Kitano:2003wn}, making 
forthcoming data from $\mu N\to eN$ competitive with $\mu\to e\gamma$.

LFV Higgs decays: A BR$(h,H,A\to\tau\mu)\sim10^{-4}$ for large $\tan\beta$
and $M_{\rm SUSY}\lsim 1$~TeV (at reach of LHC for $M_A\lsim 300$~GeV)
is compatible with limits on BR$(\tau\to\mu\gamma)$ \cite{Brignole:2003iv}.

LFV $Z$ decays: The BR$(Z\to\ell_I\ell_J)$ (dominantly chirality-preserving) 
are related but not proportional to the BR$(\ell_J\to\ell_I\gamma)$.
BR$(Z\to\tau\mu,\tau e)\lsim10^{-8}$ (at reach of GigaZ)
compatible with present limits on BR$(\tau\to\mu\gamma,e\gamma)$ are
achievable for low $\tan\beta$ \cite{Illana:2002tg,Cao:2003zv}.

Collider tests: At the LC the main signal is production and 
decay of sleptons, charginos and neutralinos. For different mSUGRA benchmark
points $\sigma(e^+e^-\to\tau^+\mu^-+2\tilde\chi^0_1)=0.05-10$~fb at
$\sqrt{s}=800$~GeV are possible and compatible with BR$(\tau\to\mu\gamma)\lsim
10^{-8}$ (foreseen at the LHC) \cite{Deppisch:2004pc}.

\section{Conclusions}

SUSY GUT models predict a misalignment of lepton and slepton mass matrices
even if one asumes a flavor-blind SUSY breaking mechanism 
(mSUGRA, GMSB, AMSB), and correlate 
leptonic and hadronic FCNC and CP. LFV and CPV in these models
should be observed in upcoming experiments.

\section{Acknowledgments}
 
J.I.I. would like to thank the conveners of the Flavor
and CP Session for their kind invitation.
Work supported by the Spanish MCYT (FPA2003-09298-C02-01), the
Junta de Andaluc{\'\i}a (FQM-101) and the European
%Human Potential Programme 
HPRN-CT-2000-00149 (Physics at colliders).

\bibliographystyle{plain}

\end{document}